\documentclass[aps,prl,twocolumn,superscriptaddress,groupedaddress]{revtex4} 
\usepackage{amssymb,bm,color,dcolumn,graphicx}

\begin{document}

\title{Onset of a skyrmion phase by chemical substitution in MnGe chiral magnet}
\author{E. Altynbaev}
\affiliation{Petersburg Nuclear Physics Institute National Research Center "Kurchatov Institute", Gatchina, 188300 St-Petersburg, Russia}
\affiliation{Vereshchagin Institute for High Pressure Physics, Russian Academy of Sciences, 142190, Troitsk, Moscow, Russia}
\author{N. Martin}
\affiliation{Laboratoire L\'eon Brillouin, CEA, CNRS, Universit\'e Paris-Saclay, CEA Saclay 91191 Gif-sur-Yvette, France}
\author{A. Heinemann}
\affiliation{Helmholtz Zentrum Geesthacht, 21502 Geesthacht, Germany}
\author{L. Fomicheva}
\affiliation{Vereshchagin Institute for High Pressure Physics, Russian Academy of Sciences, 142190, Troitsk, Moscow, Russia}
\author{A. Tsvyashchenko}
\affiliation{Vereshchagin Institute for High Pressure Physics, Russian Academy of Sciences, 142190, Troitsk, Moscow, Russia}
\author{I. Mirebeau}
\affiliation{Laboratoire L\'eon Brillouin, CEA, CNRS, Universit\'e Paris-Saclay, CEA Saclay 91191 Gif-sur-Yvette, France}
\author{S. Grigoriev}
\affiliation{Petersburg Nuclear Physics Institute National Research Center "Kurchatov Institute", Gatchina, 188300 St-Petersburg, Russia}
\affiliation{Vereshchagin Institute for High Pressure Physics, Russian Academy of Sciences, 142190, Troitsk, Moscow, Russia}
\affiliation{Faculty of Physics, Saint-Petersburg State University, 198504 Saint Petersburg, Russia}

\begin{abstract}
We study the evolution of the magnetic phase diagram of Mn$_{1-x}$Fe$_{x}$Ge alloys with concentration $x$ ($0 \leq x \leq 0.3$) by small-angle neutron scattering. We unambiguously observe the absence of a skyrmion lattice (or A-phase) in bulk MnGe and its onset under a small Mn/Fe substitution. The A-phase is there endowed with an exceptional skyrmion density, and is stabilized within a very large temperature region and a field range which scales with the Fe concentration. Our findings highlight the possibility to fine-tune properties of skyrmion lattices by means of chemical doping.
\end{abstract}

\maketitle


{\it Introduction.} The incommensurate magnetic orderings of alloys with B20 structure, such as MnSi or FeGe, have received an increasing attention in the last decade due to their peculiar magneto-transport properties. Their helical spin structure results from a competition between ferromagnetic (FM) exchange and antisymmetric Dzyaloshinskii-Moriya interaction (DMI), allowed by the lack of inversion symmetry in the crystal structure \cite{Ishikawa1976,Lebech1989,Nakanishi1980,Bak1980}. The presence of helical Bragg peaks in the direction perpendicular to the applied field $\mathbf{H}$, initially discovered in bulk MnSi single crystal by small-angle neutron scattering (SANS) \cite{Lebech1995,Grigoriev2006} was later on ascribed to a stacking of two-dimensional lattice of magnetic defects called "skyrmions" (SK) \cite{Muhlbauer2009}. The SK lattice, an hexagonal pattern with wavevector $\mathbf{k_{\rm A}} \, \bot \, \mathbf{H}$, was further observed in real space by Lorentz transmission electron microscopy (TEM) \cite{Yu2010}. Recent theories show that it results from uniaxial anisotropy, due to DMI in bulk or interfaces in layers, both reducing the effective symmetry \cite{Rybakov2015}. In the bulk, a stable SK lattice is actually only observed in a limited (H,T) region just below the ordering temperature $T_{\rm C}$, the so-called "A-phase". This suggests that chiral fluctuations, numerous around $T_{\rm C}$ \cite{Grigoriev2005,Grigoriev2011,Janoschek2013}, are needed for its stabilization together with the DMI term.

 In the B20 family, MnGe helical magnet synthesized in metastable form under high pressure and temperature \cite{Tsvyashchenko2012}, stands as an exception. MnGe orders at high temperature (T$_C$ $\sim$ 170\,K,) with much shorter helical wavelength ($\lambda_s=2\pi /k_s$ $\sim$ 2-3 nm) \cite{Makarova2012,Kanazawa2011} than MnSi or FeGe. This strongly suggests that sizable next-nearest antiferromagnetic (AFM) interactions are responsible for the helical structure \cite{Chizhikov2013,Altynbaev2016a}. MnGe also exhibits a magnetic order-disorder transition spanning a large temperature range \cite{Altynbaev2014}, and involving low-energy spin fluctuations \cite{Martin2016,Martin2019}. In MnGe, the SK lattice remains elusive \cite{Kanazawa2012} and exotic monopole defects have been further proposed \cite{Kanazawa2016}. 
 
Remarkably, {\it ab initio} calculations show that the DMI term is close to zero in MnGe, and increases under Mn/Fe substitution \cite{Gayles2015,Koretsune2018}. In Mn$_{1-x}$Fe$_{x}$Ge compounds, the ground state helical structure remains essentially similar up to x=0.35 \cite{Grigoriev2013}, but the borders of the A-phase have not yet been directly observed. Therefore, the Mn-rich Mn$_{1-x}$Fe$_{x}$Ge compounds are of a great interest to study the influence of the DMI term on the stability of the SK lattice.

\begin{figure*}[!ht]
	\includegraphics[width=0.65\textwidth]{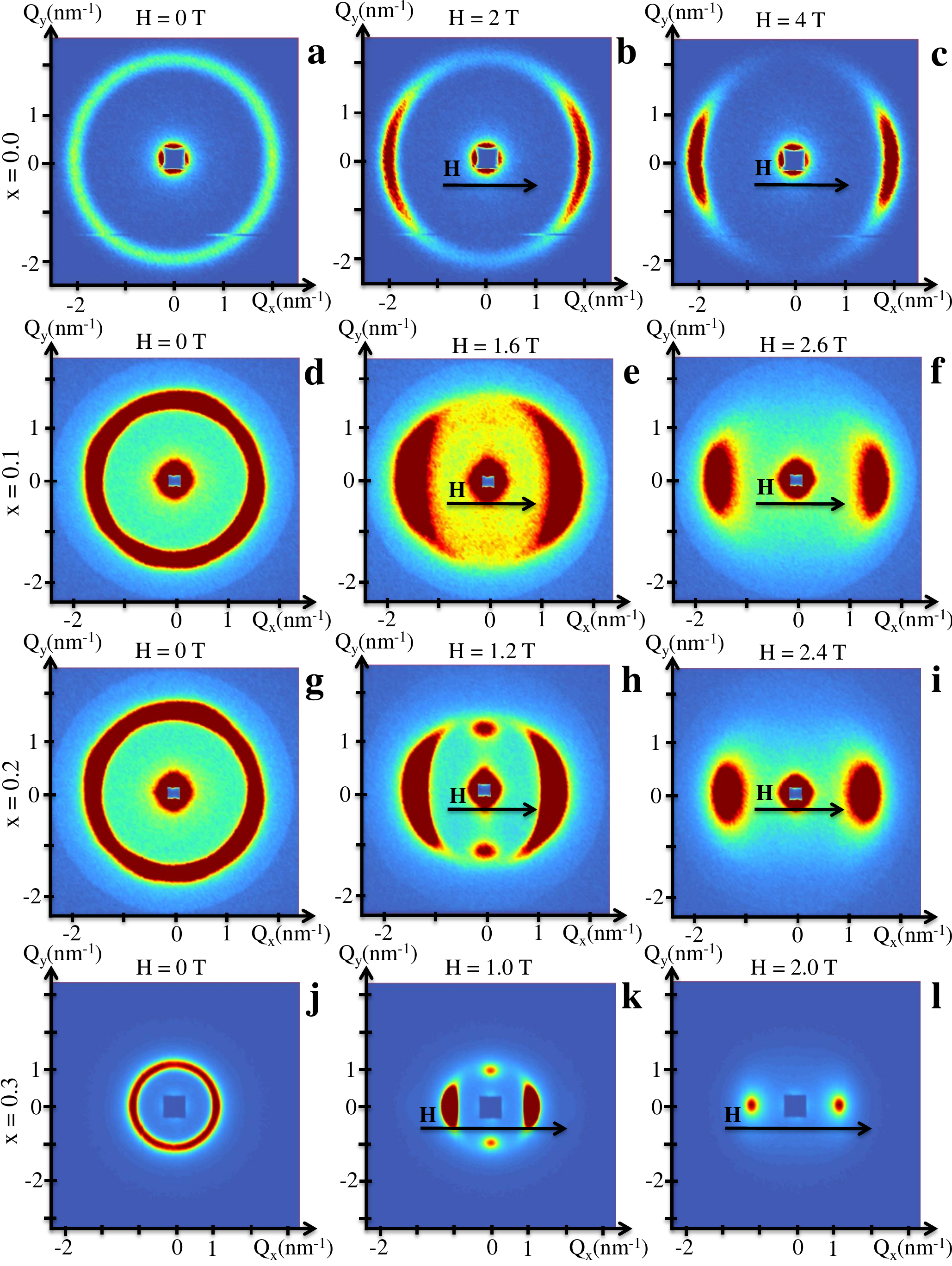}
	\caption{Small-angle scattering intensity maps, taken at different fields at $T = 100$ K on Mn$_{1-x}$Fe$_{x}$Ge compounds with $x = 0$ (a-c), $0.1$ (d-f), 0.2 (g-i) and 0.3 (j-l).}
	\label{fig:maps}
\end{figure*}

\begin{figure}[!ht]
	\includegraphics[width=0.47\textwidth]{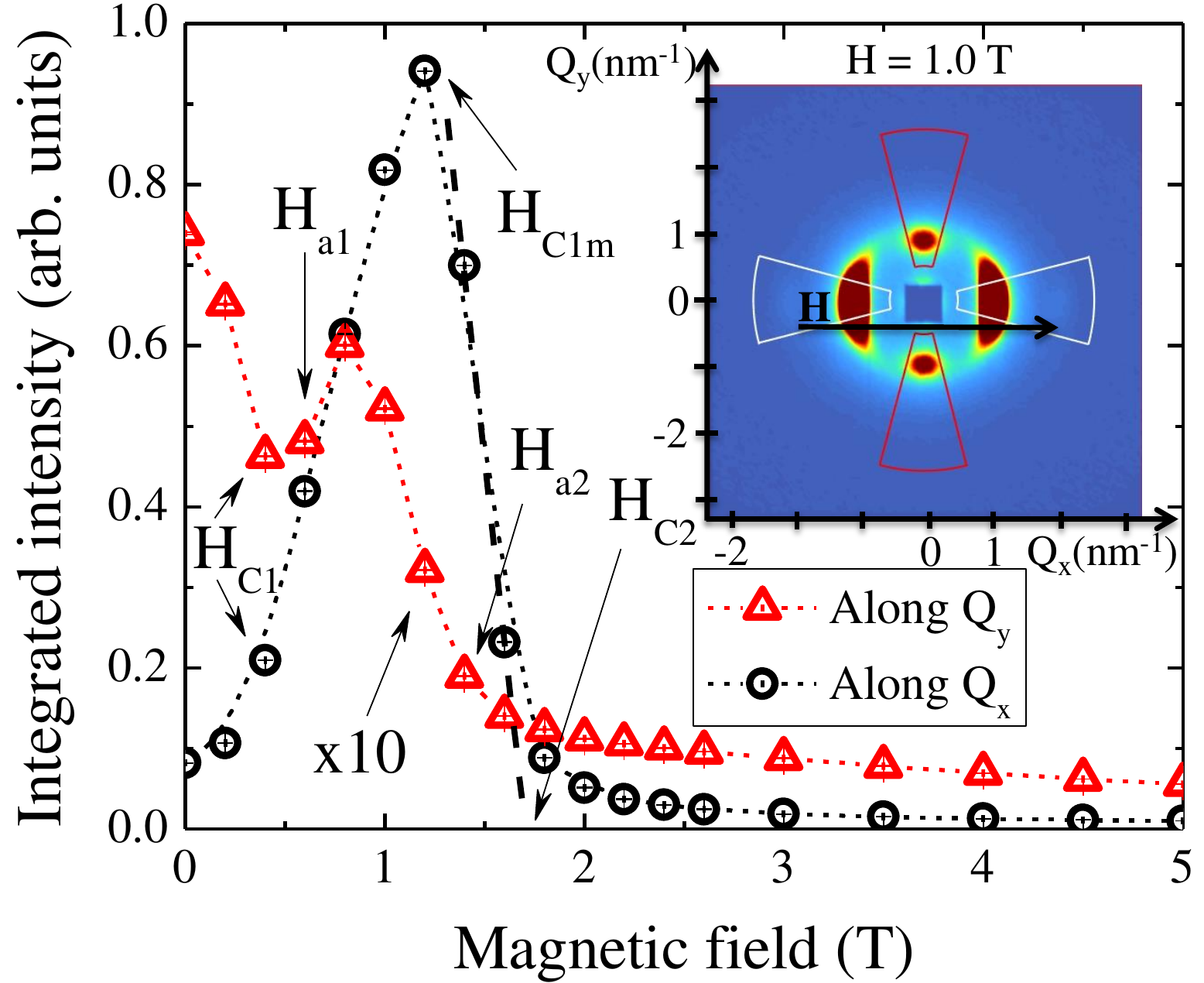}
	\caption{Neutron scattering intensity deduced from the SANS maps of Mn$_{0.7}$Fe$_{0.3}$Ge, integrated in directions along (black circles) and perpendicular (red triangles) to the external field $H$ at $T = 100$ K. A map measured at 100\,K and 0.8\,T is shown in the inset with the white (red) integration sector for the direction along (perpendicular to) the magnetic field.}
	\label{fig:hes}
\end{figure}

In this Letter, we present a comprehensive SANS study of the magnetic structure of Mn$_{1-x}$Fe$_{x}$Ge compounds with $0\leq x \leq 0.3$ under an applied magnetic field. We have used the same protocol for all samples, a well-defined procedure allowing one to determine the boundaries of the A-phase without ambiguity. We find that pure MnGe does not show any traces of a SK lattice within the explored temperature ($T \leq 200\,\text{K}$) and field ($0 \leq H \leq 9\,\text{T}$) ranges. An increase of Fe concentration, $x$, results in the appearance of the A-phase with the shortest period ever observed. This phase extends over a wide temperature range, almost independent of $x$, whereas its field range \textit{increases} with $x$. The latter scales with the calculated DMI term. Our finding agree with theoretical calculation of an almost zero DMI term in MnGe, and support the A-phase as an inherent feature of DMI helimagnets.


\begin{figure*}[!ht]
	\includegraphics[width=1.0\textwidth]{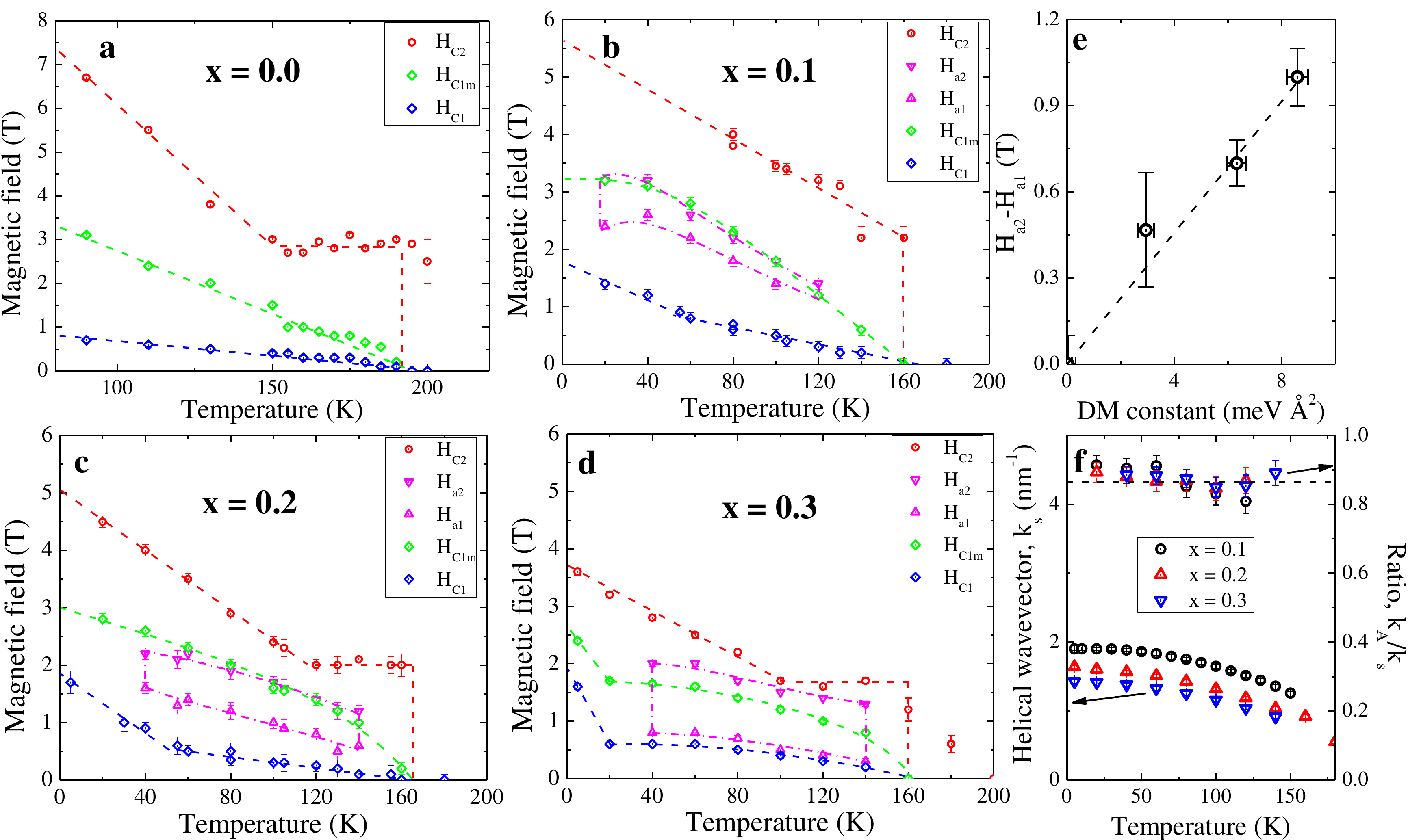}
	\caption{$(H,T)$ phase diagrams of Mn$_{1-x}$Fe$_{x}$Ge, with $x = 0.0$ (a), 0.1 (b), 0.2 (c) and 0.3 (d). (e): Field extension of the A-phase (averaged over temperature for each sample) versus the calculated DMI-constant \cite{Koretsune2018,Gayles2015,Kikuchi2016,Mankovsky2018}. Dashed line is the linear fit of the width of the A-phase versus DMI-constant plot with intercept equal to zero. (f) Temperature dependence of the helical wavevector $k_s$ (left) and ratio of the wavevector $k_{A}$ of the A-phase over $k_s$, $k_{A}/k_s$ (right), for all samples.}
	\label{fig:A-phase}
\end{figure*}

{\it Experimental results.} Polycrystalline Mn$_{1-x}$Fe$_{x}$Ge samples were previously used in Ref. \onlinecite{Altynbaev2016a}. Details on their synthesis are given in the Supplementary Material \cite{SM}. SANS experiments were carried out on the SANS-1 instrument at the Heinz Maier-Leibnitz Institute (MLZ, Garching, Germany \cite{Heinemann2015}) and on the PA20 instrument at the Laboratoire L\'eon Brillouin, (LLB, Saclay, France \cite{Chaboussant2012}), covering momentum transfers in the range $0.2 \leq Q \leq 2.7$ nm$^{-1}$. The scattered intensity was systematically measured \emph{after zero-field cooling} from $T = 300$ K (in the paramagnetic state) down to the chosen temperature, and upon a gradual increase of the magnetic field up to 9 T. 

Fig. \ref{fig:maps} shows examples of SANS maps taken at different fields at $T = 100$ K on Mn$_{1-x}$Fe$_{x}$Ge compounds. In pure MnGe (Figs. \ref{fig:maps} a-c), the isotropic ring observed in zero field transforms into a moon-like pattern oriented along the field, as expected from the evolution of the helical structure towards the conical one. However, we do not observe peaks in the direction perpendicular to the field, whatever the field and temperature up to 9\,T and 200\,K. The latter are the usual hallmark of the A-phase, within which the longitudinal magnetization is modulated in the form of a SK lattice. Strikingly, they appear in the substituted samples starting from the lowest concentration $x = 0.1$, as a weak signal superimposed on the ring structure (Fig. \ref{fig:maps} d-f). The spots from the A-phase become better-defined for the samples with $x$ = $0.2$ and $0.3$ (Fig. \ref{fig:maps} g-i and j-l, respectively). They directly demonstrate the transition of part of the samples into the A-phase and the emergence of SK lattices.

In order to determine the (H,T) phase diagrams of the studied samples, the neutron scattering intensities were integrated in the horizontal or vertical directions over the azimuthal angle of $30^{\circ}$, {\it i.e.} longitudinal or transverse to the applied field, respectively. From the field-dependence of the intensity at a given temperature, up to five characteristic fields can be deduced, as shown in Fig. \ref{fig:hes} for $x$ = 0.3 at $T = 100$ K. The critical field $H_{\rm C1}$, which we consider as the field value where the longitudinal and transverse intensities differ by more than 20\%, indicates the departure from the multi-domain helical state. The field $H_{\rm C1m}$ where the longitudinal intensity reaches its maximum, marks the end of this reorientation process and the transition of each sample grain into a single domain conical state. The critical field $H_{\rm C2}$, determined by extrapolating to zero the linear decay of the longitudinal intensity with the field increase above $H_{\rm C1m}$, marks the transition from the conical to the field-induced ferromagnetic state. The fields $H_{\rm a1}$ and $H_{\rm a2}$ are determined as the borders of the field range where the intensity in the direction, transverse to the external field, increases, indicating the SK peaks to emerge from the ring-like signal. These fields correspond to the lower and upper limit of the A-phase, respectively. Details on the accurate determination of the values of $H_{\rm a1}$ and $H_{\rm a2}$ are given in the Supplementary Material \cite{SM}. The resulting (H,T) phase diagrams are shown in Fig. \ref{fig:A-phase}a for MnGe and Fig. \ref{fig:A-phase}b-d for the substituted compounds.

In Fig. \ref{fig:A-phase}e, we plot the quantity $H_{\rm a2} - H_{\rm a1}$ which marks the field-extension of the A-phase for each concentration versus the DMI constant $D$ deduced from ab-initio calculations. The error bars on $H_{\rm a2} - H_{\rm a1}$ take into account its  temperature dependence, whereas the DMI constant is averaged over several theoretical works \cite{Gayles2015,Kikuchi2016,Mankovsky2018}. Both quantities have negligible values for pure MnGe, and are linearly increasing with concentration $x$ within error bars.

The positions of the Bragg reflections yield the periodicity of the helical structure and of the SK lattice. They can be determined for each sample as a function of field and temperature. The wavevectors of the helical structure ($k_{\rm s}$) and SK lattice ($k_{\rm A}$) are almost independent of the applied field. They slowly increase with decreasing temperature, in the same way for each sample (Fig. \ref{fig:A-phase}f). Strikingly, their ratio ($k_{\rm A}$/$k_{\rm s}$) remains almost constant, independent of the temperature and sample considered, close to the value $k_{\rm A}/k_{\rm s} \approx 0.866 \sim \sqrt{3}/2$.


{\it Discussion.} 
 In pure MnGe, the critical fields measured at low temperature $H_{\rm C1} \approx 3\,\text{T}$ and $H_{\rm C1m} \approx 5\,\text{T}$ are the highest measured in B20 compounds so far. Upon heating, they decrease to zero at $T \approx 190\,\text{K}$ (Fig. \ref{fig:A-phase}a). As a main result, we find no traces of the SK lattice when performing a careful search in the whole (H,T) range up to 9\,T and 200\,K. The critical field $H_{\rm C2}$ decreases linearly upon heating down to 3 T at $T = 150 \pm 2 \text{K}$, then saturates and remains constant up to 190\,K. As long as the traces of the helical structure persist up to 190\,K the temperature range 150\,K-190\,K likely consists of a mixed state where helical fluctuations and ferromagnetic nano-regions coexist in the sample \cite{Altynbaev2014}. 

In the substituted compounds Mn$_{1-x}$Fe$_{x}$Ge with $x = 0.1$, 0.2 and 0.3 (Fig. \ref{fig:A-phase}b-d), the temperature variation of the critical fields $H_{\rm C1}$, $H_{\rm C1m}$ and $H_{\rm C2}$ is similar to that of  MnGe. $H_{\rm C1}$ and $H_{\rm C1m}$ are almost independent of $x$ and decrease to zero at $T \approx 160$ K. The main difference with MnGe is the occurrence of a A-phase in a wide (T, H) range. It is observed for 20\,K $< T < 120$\,K  ($x=0.1$) and 40\,K $< T < 140$ K (for $x=0.2, 0.3$). The A-phase extends widely in the oriented helical-phase, between $H_{\rm C1}$ and $H_{\rm C1m}$ ($x=0.1, 0.2$), or slightly above $H_{\rm C1m}$ ($x$=0.3). The large extension of the A-phase with temperature likely result from the intrinsic instability of MnGe \cite{Deutsch2014,Altynbaev2014,Martin2016} and Mn$_{1-x}$Fe$_{x}$Ge \cite{Altynbaev2016b,Altynbaev2016a}, favoring helical fluctuations well below the ordering temperature. Strikingly, in these Mn-rich compounds where the temperature extension of the A-phase does not depend much of $x$, its field extension $H_{\rm a2} - H_{\rm a1}$ increases with $x$, in a linear way (within error bars), as does the calculated DMI constant \cite{Gayles2015,Kikuchi2016,Mankovsky2018}. The proportional increase of these two quantities (Fig. \ref{fig:A-phase}e) supports the DMI as the fundamental interaction needed to stabilize the A-phase in bulk B20 magnets.  
   
The difference between the wavevectors of the helical structure and SK lattice $k_{\rm A}/k_{\rm s} \approx 0.866$ is surprising. It means that the period of the SK is bigger than the period of the helical structure by almost 15\% whatever the concentration and temperature. We should also note that simple geometric arguments in case of the hexagonal SK lattice imply an opposite ratio, namely $k_s/k_A = \sqrt{3}/2$ \cite{Grigoriev2014}. However, it is found experimentally that $k_s = k_A$ within 1-2\%, either in bulk \cite{Lebech1995,Grigoriev2006,Ishimoto1990} or in two-dimensional \cite{Yu2010,Tonomura2012} helimagnets, with the exception of the frustrated disordered Co-Zn-Mn alloys \cite{Karube2016}. Nevertheless, the observed ratio between $k_{\rm A}$ and $k_{\rm s}$ could suggest that the SKs found within the A-phase of Mn$_{1-x}$Fe$_{x}$Ge ($x = \{0, 0.1, 0.2, 0.3\}$) are not packed in a regular hexagonal fashion. We thus speculate that the observed difference might be related to the competition between FM and AFM exchange interactions or to chemical disorder (or both). This point deserves further theoretical and experimental studies. 

The absence of a regular SK lattice in bulk MnGe suggests to reinterpret the first investigations of its (H,T) phase diagram. In the initial SANS experiments of Kanazawa {\it et al.} \cite{Kanazawa2012}, the intensity peak attributed to the A-phase was observed after the application of a large field, that could orient not only the magnetic but also the crystal domains along the field direction. Therefore an alternative scenario involving helices blocked in the hard directions could explain such intensity. Observations of the SK lattice by TEM \cite{Tanigaki2015} may be impacted by multidomain structure or surface anisotropy. On the other hand, the absence of the SK lattice in bulk MnGe is fairly natural from a theoretical viewpoint, taking into account the vanishingly small value of its DM-constant. The question that remains open concerns the origin of the large topological Hall-effect (THE). Besides SK's, other topological objects have been proposed in MnGe, such as monopole \cite{Kanazawa2016} or solitons defects \cite{Martin2019}, owing to its intrinsic instability \cite{Deutsch2014,Altynbaev2014,Martin2016}. They may provide another source for the THE.


{\it Conclusion.} We have observed the absence of a regular SK lattice or A-phase in MnGe and its onset under a small substitution of Mn for Fe. The A-phase is observed over a wide temperature range, thanks to the inherent fluctuations and metastable character of MnGe. Its field range increases linearly with the Fe concentration and calculated DMI-constant. These results emphasize that DMI and helical fluctuations are the main ingredients for the stabilization of a SK lattice in B20 magnets, and indicate a way to fine-tune the properties of dense SK lattice using controlled chemical substitution.


{\it Acknowledgements.} E.A., A.T. and S.G. are grateful to the Russian Scientific Foundation (Grant No. 17-12-01050).


\bibliographystyle{apsrev}
\bibliography{mnfege_aphase}

\end{document}


\title{Supplementary material for\\\emph{"Onset of a skyrmion phase by chemical substitution in MnGe chiral magnet"}}
\author{E. Altynbaev}
\affiliation{Petersburg Nuclear Physics Institute National Research Center "Kurchatov Institute", Gatchina, 188300 Saint-Petersburg, Russia}
\affiliation{Vereshchagin Institute for High Pressure Physics, Russian Academy of Sciences, 142190, Troitsk, Moscow, Russia}
\author{N. Martin}
\affiliation{Laboratoire L\'eon Brillouin, CEA, CNRS, Universit\'e Paris-Saclay, CEA Saclay 91191 Gif-sur-Yvette, France}
\author{A. Heinemann}
\affiliation{Helmholtz Zentrum Geesthacht, 21502 Geesthacht, Germany}
\author{L. Fomicheva}
\affiliation{Vereshchagin Institute for High Pressure Physics, Russian Academy of Sciences, 142190, Troitsk, Moscow, Russia}
\author{A. Tsvyashchenko}
\affiliation{Vereshchagin Institute for High Pressure Physics, Russian Academy of Sciences, 142190, Troitsk, Moscow, Russia}
\author{I. Mirebeau}
\affiliation{Laboratoire L\'eon Brillouin, CEA, CNRS, Universit\'e Paris-Saclay, CEA Saclay 91191 Gif-sur-Yvette, France}
\author{S. Grigoriev}
\affiliation{Petersburg Nuclear Physics Institute National Research Center "Kurchatov Institute", Gatchina, 188300 Saint-Petersburg, Russia}
\affiliation{Vereshchagin Institute for High Pressure Physics, Russian Academy of Sciences, 142190, Troitsk, Moscow, Russia}
\affiliation{Faculty of Physics, Saint-Petersburg State University, 198504 Saint-Petersburg, Russia}


\begin{abstract}
In this supplement, we provide information concerning the synthesis of the samples studied in the paper (Sec. \ref{sec:samples}). Additional details concerning the analysis of the small-angle neutron scattering data and the method used to determine the borders of the A-phase are also given (Sec. \ref{sec:sans}).
\end{abstract}

\maketitle


\section{Samples}
\label{sec:samples}

The cubic phases of Ge-based B20 alloys can only be stabilized under high pressure/high temperature conditions \cite{Tsvyashchenko2012}. The samples taken for this study have thus been synthesized under 8 GPa in a toroidal high-pressure apparatus by melting reaction with Mn, Fe and Ge at the Institute of High Pressure Physics (Troitsk, Moscow, Russia). Pellets of well-mixed powdered constituents were placed in rock-salt pipe ampoules and then directly electrically heated to T $\approx$ 1600$^\circ$C. Then, the samples were quenched to room temperature before releasing the applied pressure. The total mass of each batch is $\approx 100-150\,\text{mg}$.\\ 

As a consequence of the synthesis procedure, the samples have a metastable crystal structure and are obtained in a polycrystalline form, with crystallite sizes larger than $10-100$ $\mu$m. X-ray powder diffraction \cite{Dyadkin2014,Valkovskiy2016} confirmed the B20 structure of the samples used in the small-angle neutron scattering (SANS) experiments presented in the main text. Previous SANS studies of these samples revealed the helimagnetic ordering of the compounds at low temperatures and yielded a preliminary definition of the critical temperatures (see Refs. \onlinecite{Altynbaev2014,Altynbaev2016a,Altynbaev2016b} for more details).\\


\section{Small-angle neutron scattering data analysis}
\label{sec:sans}


\subsection{Lineshape analysis}
\label{sec:sans_data_ijkl}

\begin{figure}
	\includegraphics[width=.9\textwidth]{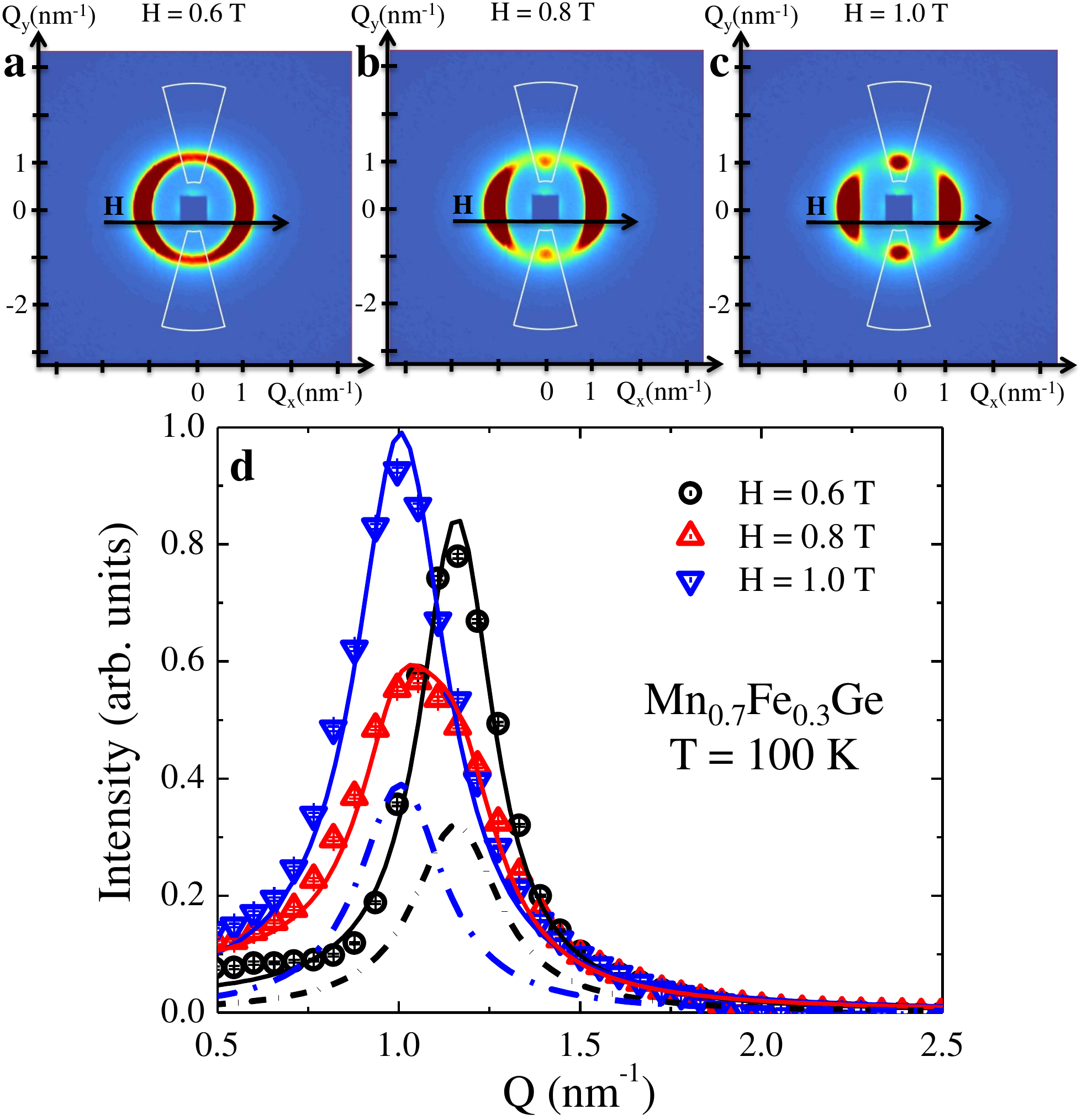}
	\caption{\textbf{(a--c)} Small-angle scattering maps, taken at different fields at $T = 100$ K for Mn$_{0.7}$Fe$_{0.3}$Ge. The white sectors were used for radial averages of the intensity. \textbf{(d)} $I \text{ vs. } Q$ plots in the direction perpendicular to the external field. Best fits of Eq. \ref{eq:1} to the data for $H = 0.6$ T (helical phase) and $H = 1.0$ T (A-phase) are shown as blue and black solid lines, respectively. They essentially consist in a single Lorentzian profile. On the other hand, the experimental curve for $H=0.8$ T --which sits on the lower edge of the A-phase-- is better described with the sum of two Lorentzian functions centered in the same positions as for the lower and higher field values (shown with black and blue dashed lines, respectively). The resulting fit curve is shown as the red solid line.}
	\label{fig:IvQ_30}
\end{figure}

In a small-angle neutron scattering (SANS) experiment, the scattered intensity is recorded using a two-dimensional position-sensitive detector. Examples of such maps are presented in Figs. \ref{fig:IvQ_30}a--c. In order to perform a quantitative analysis of the observed (temperature- and field-dependent) magnetic structures, the intensity is radially averaged and the resulting $I\,\text{vs.}\,Q$ curves are described using the following function:

\begin{equation} \label{eq:1}
I(Q) = I_{bckg} + \, A \cdot \underbrace{\frac{\kappa_s / \pi}{\kappa_s^2 + \left(Q - k_s\right)^2}}_{L_s} + \, B \cdot \underbrace{\frac{\kappa_A / \pi}{\kappa_A^2 + \left(Q - k_A\right)^2}}_{L_A} + \, C \cdot I_{abn}(Q) \quad ,
\end{equation}

where:

\begin{itemize}
	\item $I_{bckg}$ is a Q-independent background level,
	\item $A \cdot L_s$ is a Lorentzian profile centered at $Q = k_s$ with a half-width at half-maximum $\kappa_s$, which corresponds to the scattering due to the (incompletely reoriented) spin spirals,
	\item $B \cdot L_A$ is a Lorentzian profile centered at $Q = k_A$ with a half-width at half-maximum $\kappa_A$, which corresponds to the scattering due to the SK lattices stabilized within the A-phase,
	\item $C \cdot I_{abn}$ is a smeared Heaviside function centered at $Q = k_s$, which describes phenomenologically the inelastic scattering denoted as "abnormal" in Ref. \onlinecite{Altynbaev2014}.  
\end{itemize}

As a general trend, the Lorentzian widths and positions are found to be field-independent. In what follows, the parameters $\kappa_{s,A}$ and $k_{s,A}$ are thus kept constant for the analysis of the fixed-temperature field scans.

\subsection{Determination of the A-phase borders}
\label{sec:sans_data_30}

\begin{figure}
	\includegraphics[width=.9\textwidth]{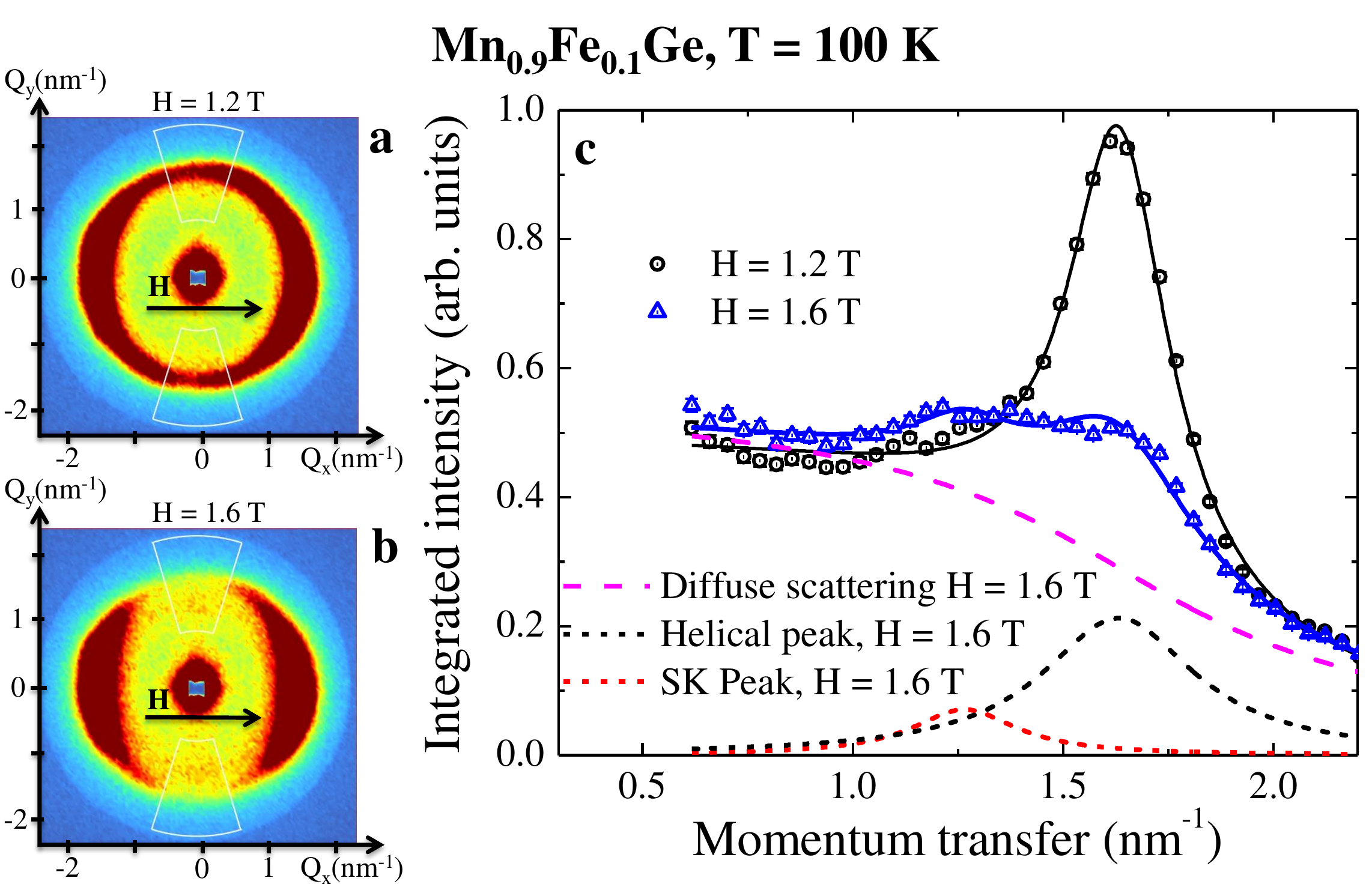}
	\caption{SANS maps, taken at $T = 100$ K for Mn$_{0.9}$Fe$_{0.1}$Ge at fields $H=1.2$ T \textbf{(a)} and $H=1.6$ T \textbf{(b)}. The white sectors in \textbf{(a--b)} have been used for performing the radial averages. \textbf{(c)} $I\,\text{vs.}\,Q$ plots in the direction perpendicular to the external field. Solid black and blue lines are fits with Eq. \ref{eq:1} of the data collected at applied magnetic fields of $H = 1.2$ T and 1.6 T, respectively. At the field of 1.2 T, the experimental profile is well-described using a single Lorentzian, sitting on top of a diffuse signal associated with the "abnormal" scattering of Eq. \ref{eq:1} (magenta dashed line). On the other hand, the description of the data at $H = 1.6$ T requires an additional Lorentzian profile (red dashed line), centered at a Q value that is smaller than the one corresponding to the helical ordering (black dashed line).}
	\label{fig:IvQ_10}
\end{figure}

The critical fields $H_{\rm a1}$ and $H_{\rm a2}$ are defined as the lower and upper borders of the A-phase, respectively (see, {\it e.g.}, Fig. 3 of main text). They were accurately determined by analyzing the $I\,\text{vs.}\,Q$ curves, obtained after radial integration of the scattered intensity in the direction perpendicular to the external field (see sectors in Fig. \ref{fig:IvQ_30}a--c). Examples of such plots are given in Fig. \ref{fig:IvQ_30}d for Mn$_{0.7}$Fe$_{0.3}$Ge.\\

At fields smaller than $H_{a1}$, the reflection coming from incompletely reoriented helical structure is always described using a single Lorentzian profile $L_s$, setting $B = 0$ in Eq. \ref{eq:1}. With field increase, the scattering peak broadens and its center of gravity shifts to lower values of momentum transfer. As illustrated in Fig. \ref{fig:IvQ_30}d, it is actually best described using two Lorentzian profiles.\\

\begin{figure}
	\includegraphics[width=\textwidth]{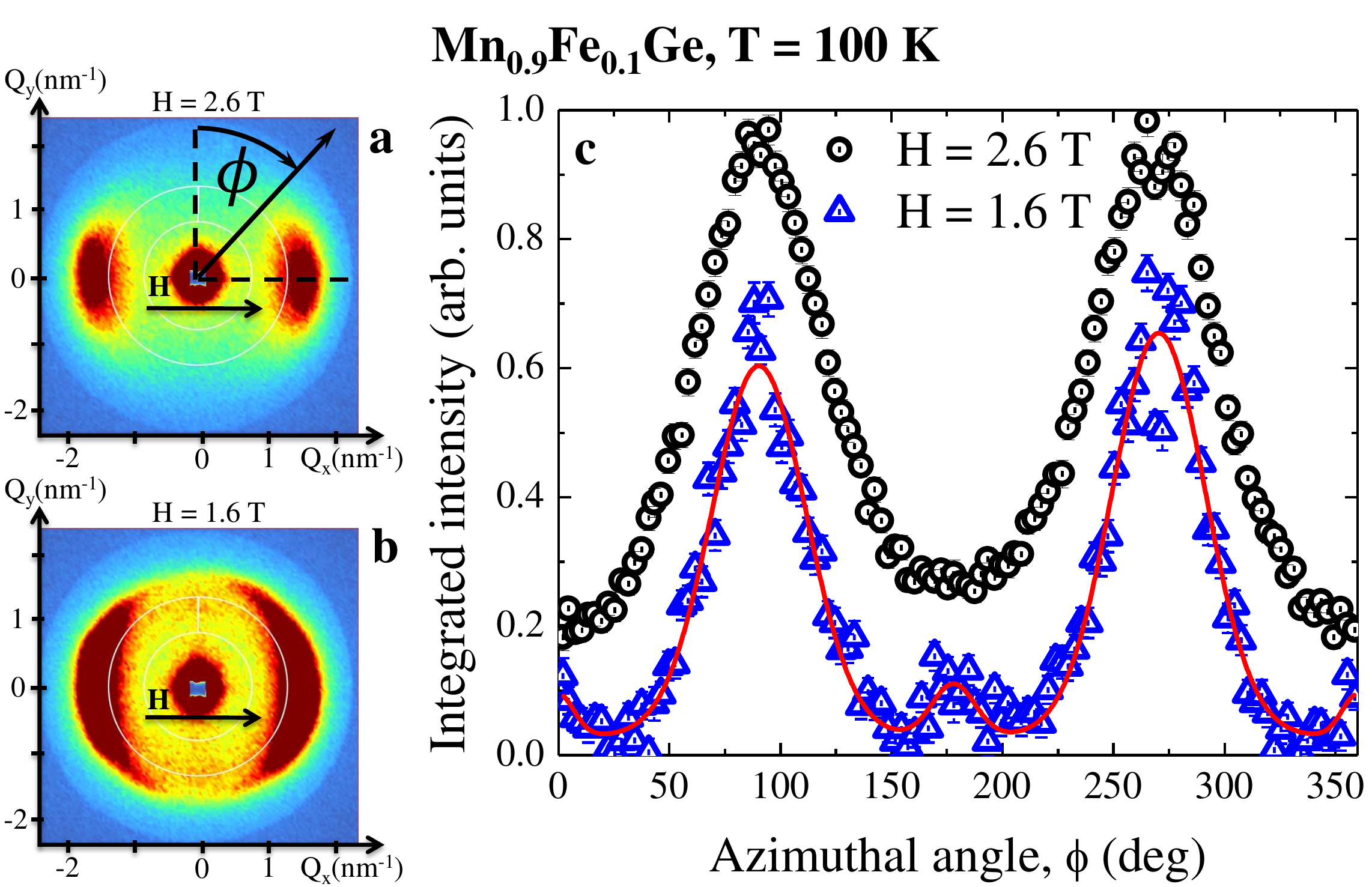}
	\caption{Small-angle scattering intensity maps, taken at $T = 100$ K for Mn$_{0.9}$Fe$_{0.1}$Ge at fields $H=2.6$ T \textbf{(a)} and $H=1.6$ T \textbf{(b)} and the corresponding $I \text{ vs. } \phi$ plots \textbf{(c)}. The sectors that have been taken for the averaging of the intensity are marked with white colour on SANS maps \textbf{(a--b)}. The value of the isotropic intensity was chosen as 0 and 0.2 for field values $H = 1.6$ T and $H = 2.6$ T, respectively, for better visibility. Red solid line is the guide for the eyes that shows peaks of the intensity for azimuthal angles equal to 0, 90, 180 and 270 degrees for experimental curve taken at $H=1.6$ T.}
	\label{fig:IvPhi_10}
\end{figure}

This suggests the emergence of an additional magnetic phase, coexisting with the conical state but showing a different periodicity. In analogy with the vast majority of cubic chiral magnet, we treat this "extra" intensity as the signature of the A-phase, populated with magnetic SKs. This is justified owing to the selection rule for magnetic neutron scattering, which dictates that only the magnetic moment component which is perpendicular to the scattering vector contributes to the scattered intensity. Here, it implies that the additional intensity reflects the spatial modulation of the longitudinal magnetization ({\it i.e.}, the component oriented along the applied field).\\
 
With further increase of the magnetic field, the first Lorentzian disappears completely ($A = 0$ in Eq. \ref{eq:1}) while the second one solely remains ($B \neq 0$ in Eq. \ref{eq:1}) up to $H_{\rm a2}$, thereby defining the upper border of the A-phase.
 


\subsection{The Mn$_{0.9}$Fe$_{0.1}$Ge case}
\label{sec:sans_data_10}

While the signal associated with the A-phase is clearly seen on the SANS maps for the Mn$_{0.8}$Fe$_{0.2}$Ge and Mn$_{0.7}$Fe$_{0.3}$Ge samples (see Fig. \ref{fig:IvQ_30} of this supplement and Figs. 1,2 of main text), it is much weaker in the case of Mn$_{0.9}$Fe$_{0.1}$Ge. However, applying the analysis strategy described above allows retrieving the (H,T) borders of the A-phase in this particular case. This fact is illustrated in Fig. \ref{fig:IvQ_10}, where a doubled peak is evidenced in the intermediate field range. A fit of Eq. \ref{eq:1} to the data indeed reveals two distinct periodicities, similar to the example given in Sec. \ref{sec:sans_data_30}.\\

%

\begin{figure}
	\includegraphics[width=0.6\textwidth]{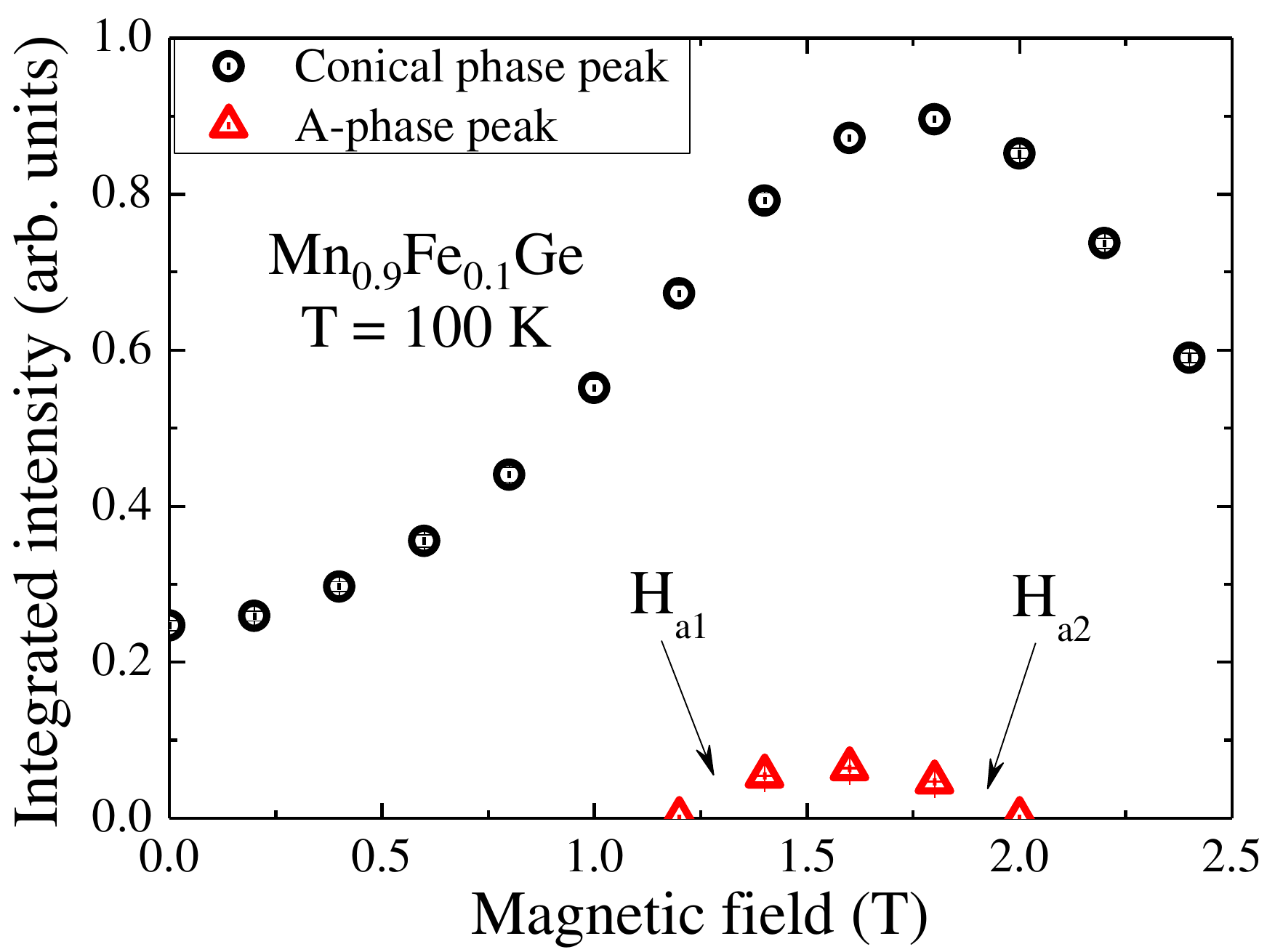}
	\caption{Field-dependence of the integrated intensities of the peaks from the helical structure (black circles) and from the A-phase (red triangles) for Mn$_{0.9}$Fe$_{0.1}$Ge.}
	\label{fig:IvsH_all}
\end{figure}

As a cross-check for the existence of a A-phase signal in Mn$_{0.9}$Fe$_{0.1}$Ge, it is also interesting to consider the azimuthal dependence of the intensity $I\,\text{vs.}\,\phi$ (Fig. \ref{fig:IvPhi_10}c). In the purely conical state ({\it e.g.}, $H = 2.6\,\text{T}$ in Fig. \ref{fig:IvPhi_10}c), the latter is composed of two peaks, centered around $\phi = 0$ and $180^{\circ}$, {\it i.e.} parallel and antiparallel to the applied field. On the other, for fields where a double peak is observed in the $I\,\text{vs.}\,Q$ plots, some additionnal intensity appears at angles $\phi = 90$ and $270^{\circ}$, {\it i.e.} perpendicular to the applied field ({\it e.g.}, $H = 1.6\,\text{T}$ in Fig. \ref{fig:IvPhi_10}c).\\

\subsection{Building the (H,T) phase diagrams}
\label{sec:sans_data_mnop}

In order to render the (H,T) phase diagrams presented in Fig. 3 of main text for all studied compositions, the field evolutions of the the $I\,\text{vs.}\,Q$ curves are considered for both parallel and perpendicular directions with respect to the applied magnetic field. This allows obtaining the critical fields attributed to the conical (along the field) and SK (perpendicular to the field) phases, by plotting the H-dependence of the fit parameters $A$ and $B$ of Eq. \ref{eq:1} (Fig. \ref{fig:IvsH_all}). Namely, the region of existence of the A-phase corresponds to the field range within which $B \neq 0$. On the other hand, the maximum of $A$ marks the first (conical) critical field $H_{\rm C1}$, while $H_{\rm C2}$ is defined through a linear extrapolation of $A \rightarrow 0$ at the largest fields.\\

\bibliographystyle{apsrev}
\bibliography{mnfege_aphase}